\documentclass[12pt]{iopart}
\usepackage{float}
\usepackage[dvips]{graphicx}%kepek beillesztesehez, rajzolashoz
\begin{document}
\title{Collision of water wave solitons}

\author{N.~Fenyvesi and
G.~Bene}
\address{
Institute of Physics, Lor\'and E\"otv\"os  University\\P\'azm\'any P\'eter s\'et\'any 1/A H-1117 Budapest, Hungary
}
\eads{\mailto{nor.fenyvesi@gmail.com}, \mailto{bene@arpad.elte.hu} }

%%\date{}
%%\maketitle

\begin{abstract}
A classification of the time evolution of the two-soliton solutions of the Boussinesq equation is given, based on the number of extrema of the wave. For solitons moving in the same directions, three different scenarions are found, while it is shown that only one of these scenarios exists in case of oppositely moving solitons. 
\end{abstract}
\section{Introduction}

Interest in environmental flows\cite{Carrier}-\cite{Voronovich} both utilize and motivate theoretical work in nonlinear wave phenomena. Among these are solitons\cite{drazin}-\cite{segurmegint}, which appear
in many different situations and have applications in several other
branches of physics\cite{Wu}, \cite{Mahmood}. Although solitons are well known and much studied,
we believe that a simple classification scheme of the evolving wave shapes of two colliding water wave solitons may not be in vain.   

Weakly nonlinear waves in a shallow channel may be well described in terms of the Boussinesq equation. It allows propagation in both directions.\footnote{Strictly speaking, the derivation of that equation does not allow counterpropagating waves.}  The small parameters of the problem are the ratio of the wave amplitude to the water depth ($\epsilon=a/h$) and the ratio of the water depth to the wave length ($\delta=(h/l)^2$), these are assumed to be comparable. The smaller these parameters are, the better the approximation will be. Therefore, any phenomenon found when solving the   Boussinesq equation corresponds to observable effects if $\epsilon$ and $\delta$ are sufficiently small.

It is well known that the    Boussinesq equation is fully integrable and has soliton solutions. Especially, when two solitons have collided, their individual properties are preserved and completely restored when they spatially separate. The change of the wave's shape during collisions is interesting and, as it turns out, follows three possible scenarios. Especially, it is intriguing, what happens if solitons with very similar parameters collide, since it is well known, that the one-soliton solution is unique (up to a scale transformation), hence, two identical, albeit spatially separated solitons do not form an exact solution.

In the present paper we classify the possible wave shape changes during the
collision of two Boussinesq solitons. 

\section{General properties of the two-soliton solution of the Boussinesq equation}

The Boussinesq equation in dimensionless form is given by
\begin{eqnarray} 
\eta_{tt}-\eta_{xx}-3\left(\eta^2\right)_{xx}- \eta_{xxxx}=0\;.\label{bq1}
\end{eqnarray}

Two-soliton solutions are generated by the Zakharov-Shabat scheme \cite{drazin}, and may be written as
\begin{eqnarray} 
\eta=-4\frac{\partial}{\partial x}\left\{\frac{k_1(1+q)+k_2(1+p)-(k_1+k_2)a}{(1+q)(1+p)-a}\right\}\;,\label{bq2}
\end{eqnarray}
where the new variables $p$ and $q$ are defined by 
\begin{eqnarray}
p={\rm e}^{2k_1 x-2\omega_1 t}\;,\quad q={\rm e}^{2k_2 x-2\omega_2 t}
\label{bq3a}
\end{eqnarray}
and the parameter $a$ is given by
\begin{eqnarray}
a=\frac{4k_1k_2}{(k_1+k_2)^2+\frac{1}{12}\left(\frac{\omega_1}{k_1}-\frac{\omega_2}{k_2}\right)^2}\;.
\label{bq3b}
\end{eqnarray}
In these expressions $k_1$ and $k_2$ are arbitrary positive parameters characterizing the individual solitons and 
\begin{eqnarray}
\omega_j=\pm k_j\sqrt{1+4k_j^2}\;,\quad (j=1,2)\label{bq3}
\end{eqnarray}
with positive sign for a soliton propagating from left to right, and negative otherwise. Note that if $\omega_1\ne \omega_2$,
\begin{eqnarray}
0<a<1\;.\label{bq3c}
\end{eqnarray}
On the other hand, if $\omega_1= \omega_2$, we also have $k_1=k_2$ that corresponds to a one-soliton solution, not being of interest here.
%%ide jönne a t -> +- infty eset rövid diszkussziója

An important observation is that in the extreme cases when $k_1,\; k_2\ll 1/2$ or $k_1,\; k_2\gg 1/2$ the parameter $a$ depends only on the ratio $$\kappa=\frac{k_2}{k_1}\;.$$
Explicitly, we get for solitons moving in the same direction
\begin{eqnarray}
a=\left\{\begin{array}{cl}
\frac{4\kappa}{(1+\kappa)^2+\frac{1}{3}(1- \kappa)^2}&{\rm if }\quad k_1,k_2\gg \frac{1}{2}\\
\frac{4\kappa}{(1+\kappa)^2}&{\rm if }\quad k_1,k_2\ll \frac{1}{2}
\end{array}\right.
  \label{bq3d}
\end{eqnarray}
and for solitons moving oppositely
\begin{eqnarray}
a=\left\{\begin{array}{cl}
\frac{4\kappa}{(1+\kappa)^2+\frac{1}{3}(1+ \kappa)^2}&{\rm if }\quad k_1,k_2\gg \frac{1}{2}\\
0&{\rm if }\quad k_1,k_2\ll \frac{1}{2}
\end{array}\right.
  \label{bq3e}
\end{eqnarray}

 For the spatial derivative of the wave we have
\begin{eqnarray} 
\frac{\partial \eta}{\partial x}=\frac{8}{\left((1+q)(1+p)-a\right)^3}\sum_{n=1}^5\sum_{j=0}^n c_{nj}p^jq^{n-j}\;,\label{bq4}
\end{eqnarray}
where the coefficients are given by
\begin{eqnarray}
c_{10}&=&-k_2^3(a-1)^2\nonumber\\
c_{11}&=&-k_1^3(a-1)^2\nonumber\\
c_{20}&=&-k_2^3(a-1)\nonumber\\
c_{21}&=&-(k_1+k_2)\left[a(k_1+k_2)^2-3k_1^2+3k_1k_2-3k_2^2\right](a-1)\nonumber\\
c_{22}&=&-k_1^3(a-1)\nonumber\\
%%c_{30}&=&0\\
c_{31}&=&(k_1-k_2)\left[a(2k_1^2+5k_1k_2+2k_2^2)-3k_1^2-3k_1k_2-3k_2^2\right]\label{bq5}\\
c_{32}&=&(k_2-k_1)\left[a(2k_1^2+5k_1k_2+2k_2^2)-3k_1^2-3k_1k_2-3k_2^2\right]\nonumber\\
%%c_{33}&=&0\\
%%c_{40}&=&0\\
c_{41}&=&-k_1^3\nonumber\\
c_{42}&=&-(k_1+k_2)\left[a(k_1+k_2)^2-3k_1^2+3k_1k_2-3k_2^2\right]\nonumber\\
c_{43}&=&-k_2^3\nonumber\\
%%c_{44}&=&0\\
%%c_{50}&=&0\\
%%c_{51}&=&0\\
c_{52}&=&k_1^3\nonumber\\
c_{53}&=&k_2^3\nonumber
%%c_{54}&=&0\\
%%c_{55}&=&0
\end{eqnarray}
All the other coefficients are zero. Again, in the extremes $k_1,\; k_2\ll 1/2$ or $k_1,\; k_2\gg 1/2$ a common factor $k_1^3$ can be pulled out of all the coefficients and the rest depends only on the ratio $k_2/k_1$. Accordingly, for both very small and very large wave numbers the behavior of the solitons is determined by this ratio, up to a scaling.   

For minima and maxima of the amplitude $\eta(x,t)$ at a given time $t$ we have
\begin{eqnarray} 
\sum_{n=1}^5\sum_{j=0}^n c_{nj}p^jq^{n-j}=0\;.\label{bq6}
\end{eqnarray}

Additionally, due to the definitions (\ref{bq3}) we have
\begin{eqnarray} 
q^{k_1}p^{-k_2}={\rm e}^{2(k_2\omega_1-k_1\omega_2)t}\;.\label{bq7}
\end{eqnarray}

\begin{figure}[H]                                                                  
\begin{center}                                                                  
\includegraphics[height=5cm]{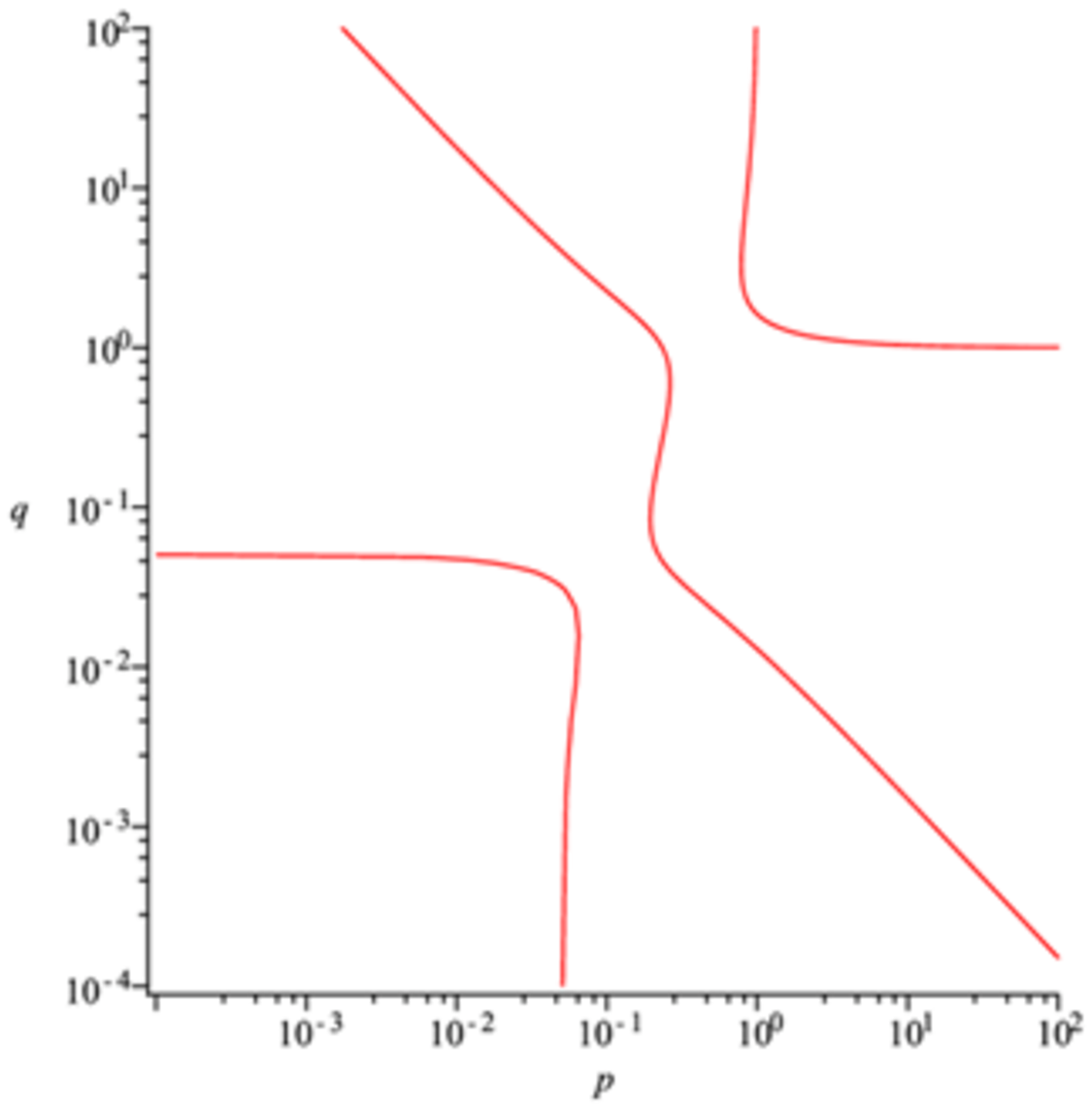}
\includegraphics[height=5cm]{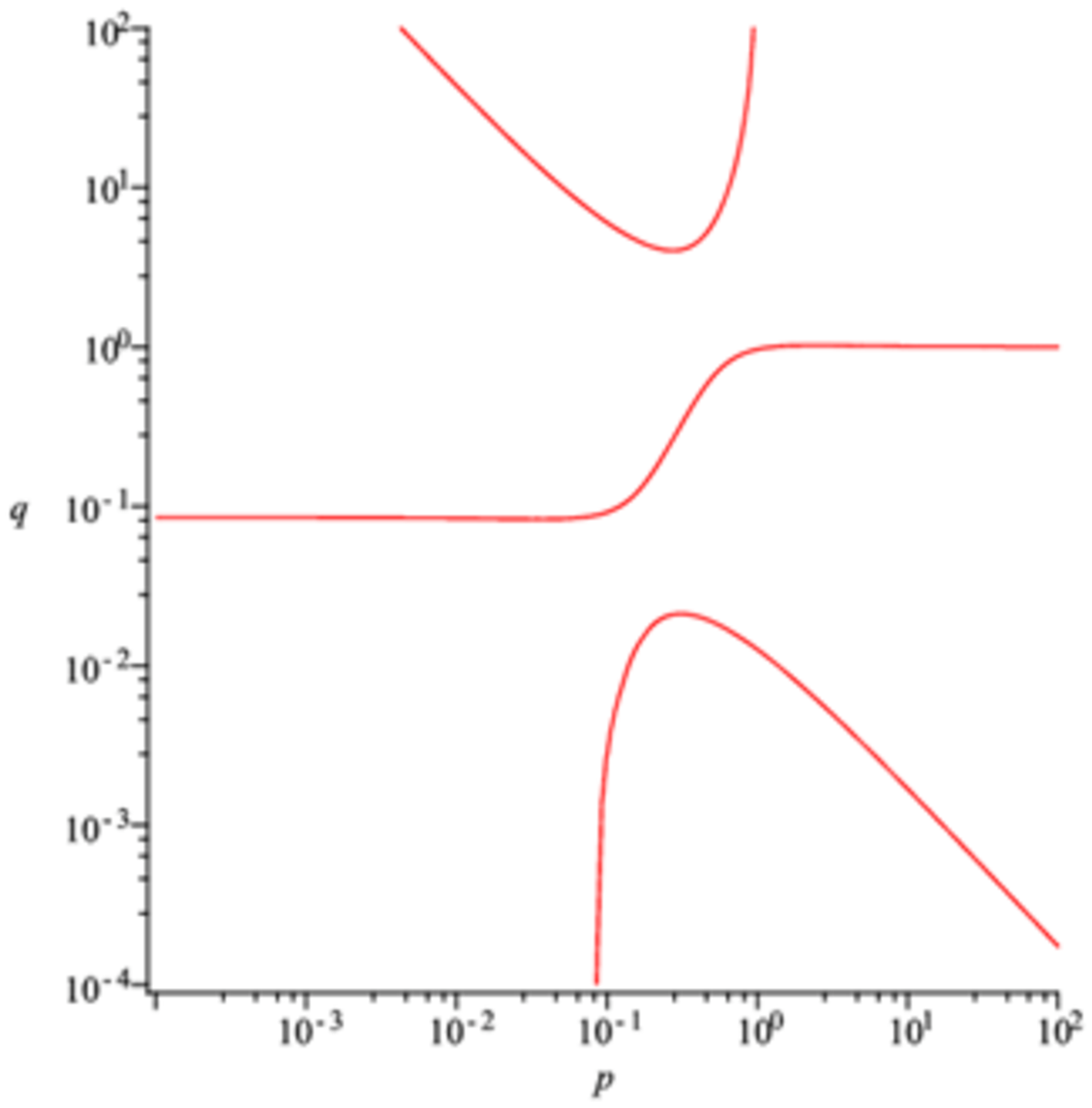}      %%[height=8cm, width=10cm]         
\caption{The zeros of Eq.(\ref{bq6}) for solitons moving in the same
  direction. Left panel: Type I. case (at $k_1=1$,
  $k_2=1.5$), right panel: Type II. case (at $k_1=1$, $k_2=1.7$). \label{f1} }\end{center}          
\end{figure}

\begin{figure}[H]                                                                 \begin{center}                                                                  \includegraphics[height=6cm]{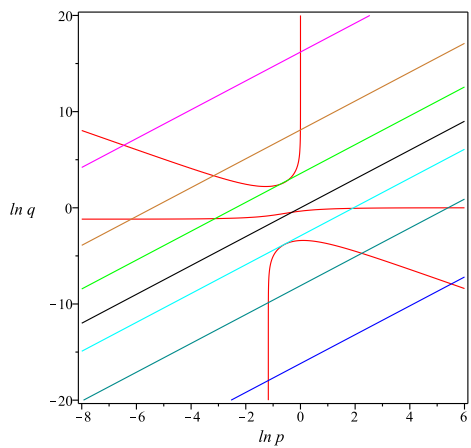}      %%[height=8cm, width=10cm]         
\caption{Type II.b topology for intersections of the curve (\ref{bq6}) and (\ref{bq7}) at parameters $k_1=1.0$ and $k_2=1.5$ for solitons moving in opposite directions. Different colors of the straight lines correspond to different time instants.            \label{f11a}}      \end{center}          
\end{figure}

Thus, extrema are given by the intersections of the curves (\ref{bq6}) and (\ref{bq7}). The graph of Eq.(\ref{bq7}) is a power function with a time-dependent coefficient. In fact, time appears only here. 
As for the graph of Eq.(\ref{bq6}), we may prove some general properties on the basis of Eqs.(\ref{bq5}), namely
\begin{enumerate} 
\item Exchanging $k_1$ with $k_2$ and $\omega_1$ with $ \omega_2$ is equivalent with exchanging $p$ with $q$. This follows directly from Eqs.(\ref{bq2})-(\ref{bq3b}).
\item The curves can have at most three intersections with a line $p=const.$ or $q=const.$ Indeed, for a fixed $p$ Eq.(\ref{bq6}) is a third order polynomial of $q$ and vica versa.
\item The transformation 
\begin{eqnarray} 
p\rightarrow (1-a)\frac{1}{p}\;,\quad q\rightarrow (1-a)\frac{1}{q}\label{id1}
\end{eqnarray}
leaves Eq.(\ref{bq6}) invariant. \hfill\break Indeed, direct substitution shows that
under transformation (\ref{id1}) the expression \hfill\break $\sum_{n=1}^5\sum_{j=0}^n c_{nj}p^jq^{n-j}$ goes over into 
\begin{eqnarray}
-\frac{(1-a)^3}{p^3q^3}\sum_{n=1}^5\sum_{j=0}^n c_{nj}p^jq^{n-j}\;.
\end{eqnarray} 

In view of the definition of $p$ and $q$, the symmetry (\ref{id1}) means that the two-soliton solution is invariant with respect to a simultaneous spatial and temporal reflection, with respect to suitably chosen origins. Explicitly, transformation (\ref{id1}) is equivalent with
\begin{eqnarray} 
x\rightarrow 2x_0-x\;,\quad t\rightarrow 2t_0 -t\;,\label{id2}
\end{eqnarray} 
where
\begin{eqnarray} 
x_0=\frac{1}{4}\frac{\omega_2-\omega_1}{k_1\omega_2-k_2\omega_1}\ln(1-a)\;,\nonumber\\
t_0=\frac{1}{4}\frac{k_2-k_1}{k_1\omega_2-k_2\omega_1}\ln(1-a)\;.\label{id3}
\end{eqnarray} 
The symmetry manifests itself in the log-log plots Fig.\ref{f1}, Fig.\ref{f8a},
Fig.\ref{f9a}, Fig.\ref{f10a}, Fig.\ref{f11a} as an inversion symmetry with
respect to the symmetry point
\begin{eqnarray}
p=q=\sqrt{1-a}\;.\label{sp} 
\end{eqnarray}

\item Although $p=0$, $q=0$ satisfies Eq.(\ref{bq6}), in the first quadrant no curve starts from the origin. This is because for small $p$ and $q$ the first order terms dominate, and (for positive $p$ and $q$) they are both negative. 
%\item Direct substitution shows that both $p=1-a$, $q=0$ and $p=0$, $q=1-a$ belong to the curve.
\item For $p\rightarrow \infty$ two asymptotes exist, namely, one at $q=0$ and 
another one at $q=1$. Similarly, for $q\rightarrow \infty$ we have asymptotes
at $p=0$ and $p=1$. This result can easily be obtained since e.g. for large values of $p$ the terms containing the highest (third) power of $p$ dominate, i.e., 
\begin{eqnarray} 
k_2^3p^{3}q^{2}-k_2^3p^{3}q=0\;.\label{bq8}
\end{eqnarray}
\item Near the above asymptotes the curve may be approximated by
\begin{eqnarray}
q&=&\frac{k_1^3}{k_2^3}(1-a)\frac{1}{p}\quad {\rm if}\quad p\rightarrow \infty\;,\;q\rightarrow 0\label{asq0}\\
1-q&=&\frac{k_1^3}{k_2^3}\left[8-a\left(4+6\frac{k_2}{k_1}-\left(\frac{k_2}{k_1}\right)^3\right)\right]\frac{1}{p}\quad {\rm if}\quad p\rightarrow \infty\;,\;q\rightarrow 1\label{asq1}\\
p&=&\frac{k_2^3}{k_1^3}(1-a)\frac{1}{q}\quad {\rm if}\quad q\rightarrow \infty\;,\;p\rightarrow 0\label{asp0}\\
1-p&=&\frac{k_2^3}{k_1^3}\left[8-a\left(4+6\frac{k_1}{k_2}-\left(\frac{k_1}{k_2}\right)^3\right)\right]\frac{1}{q}\quad {\rm if}\quad q\rightarrow \infty\;,\;p\rightarrow 1\label{asp1}
\end{eqnarray}
This can be readily shown by taking into account the next-to-highest (second) power of the large variable. 

%%This result means that the curve for large $p$ or $q$ lies within the strip bordered by the parallel asymptotes.

The right hand sides of Eqs.(\ref{asq0}), (\ref{asp0}) are positive, thus the curve approaches the asymptote from above and from the right, respectively.\footnote{Hence they lie in the first quadrant of the $(p,q)$ coordinate system.}  
The same is true for Eqs.(\ref{asq1}), (\ref{asp1}) in case of oppositely moving solitons ($\omega_1\omega_2<0$). In contrast, for solitons moving in the same direction ($\omega_1\omega_2>0$) the right hand sides of Eqs.(\ref{asq1}), (\ref{asp1}) can be both negative or positive, depending on the parameters $k_1$, $k_2$. %The corresponding ranges are shown on Figs.() and (). Note that the linear-looking borders are not exactly linear.  

\end{enumerate} 

\noindent The above properties allow us to explain the possible topologies of the curve (\ref{bq6}).

Under transformation (\ref{id1}) the point $p=1$, $q=\infty$  goes over into $p=1-a$, $q=0$, similarly, the point $p=\infty$, $q=1$ goes over into $p=0$, $q=1-a$. Finally,  the point $p=0$, $q=\infty$ goes over into $p=\infty$, $q=0$. This allows only three possible topologies, 
%as shown in Figs. (\ref{f1})-(\ref{f3}), 
according to the fact that the point $p=1-a$, $q=0$ should be continuously
connected with either $p=0$, $q=1-a$ (Fig.\ref{f1} left panel), or $p=\infty$,
$q=0$ (Fig.\ref{f1} right panel), or $p=1$, $q=\infty$. This last possibility
is visualized again by the right panel of Fig.\ref{f1} if $p$ is exchanged
with $q$, which, according to property 1 above, is equivalent with exchanging
$k_1$ with $k_2$. Since these curves are intersections of a smooth surface
with a plain, other possibilities are ruled out. We shall call the topology
shown in the left panel of Fig.\ref{f1} the Type I. case, and the topology
shown in the right panel of Fig.\ref{f1} the Type II. case. The topology
obtained from the Type II. case via exchanging the axes will be called the
Type $\overline{\rm II}$. case. Note that the same transformation does not
change the topology Type I.

While all the three situations do occur for solitons moving in the same
direction, the Type I. case never occurs for oppositely moving
solitons. This can be shown by considering the intersections of the curve
(\ref{bq6}) with the $p=q$ line, i.e., the zeros of 
\begin{eqnarray}
\xi\left[\xi^2-(1-a)\right]\left[\xi^2+b\xi+(1-a)\right]\;,\label{bq6s}
\end{eqnarray}  
where $\xi=p=q$ and
\begin{eqnarray}
b=2-a\frac{(k_1+k_2)^3}{k_1^3+k_2^3}\;.\label{bq6s1}
\end{eqnarray}
Polynomial (\ref{bq6s}) always has zeros at $\xi=0$ and $\xi=\pm
\sqrt{1-a}$. The discriminant of the last quadratic factor is
\begin{eqnarray}
D=b^2-4(1-a)=\frac{a(k_1+k_2)^6}{(k_1^3+k_2^3)^2}\left[a-\frac{12k_1k_2(k_1^3+k_2^3)}{(k_1+k_2)^5}\right]\;.\label{disc}
\end{eqnarray}   
It is a simple exercise to show that the last factor on the right hand side of
Eq.(\ref{disc}) is always negative in the case of oppositely moving solitons
(cf. Eq.(\ref{bq3b})). Hence, in that case only a single positive root exists,
while the case Type I. shown in the left panel of Fig.\ref{f1} requires
three positive roots.

\begin{figure}[H]                                                                  
\begin{center}

\includegraphics[height=5cm, width=5cm]{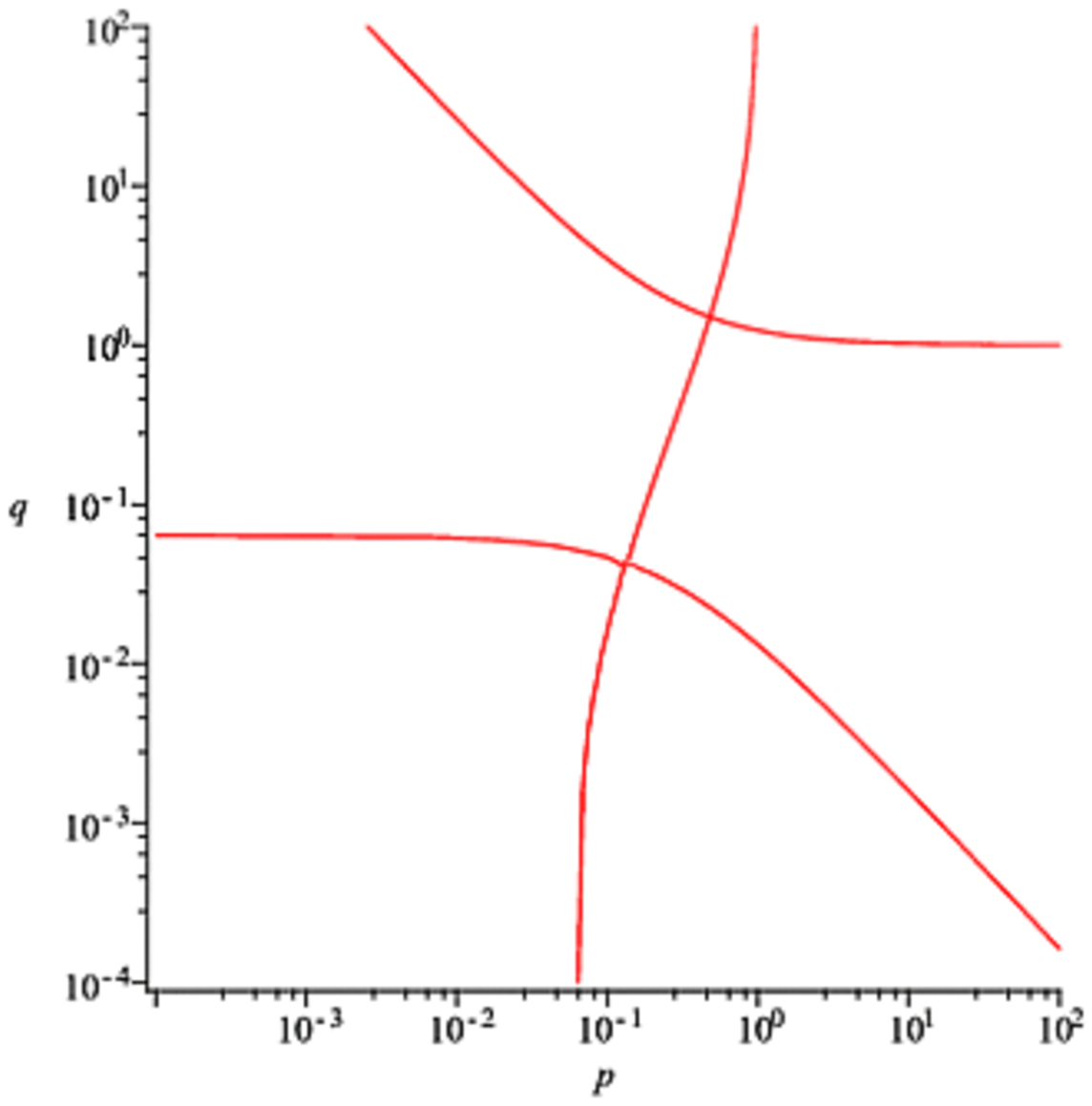}
\includegraphics[height=5cm]{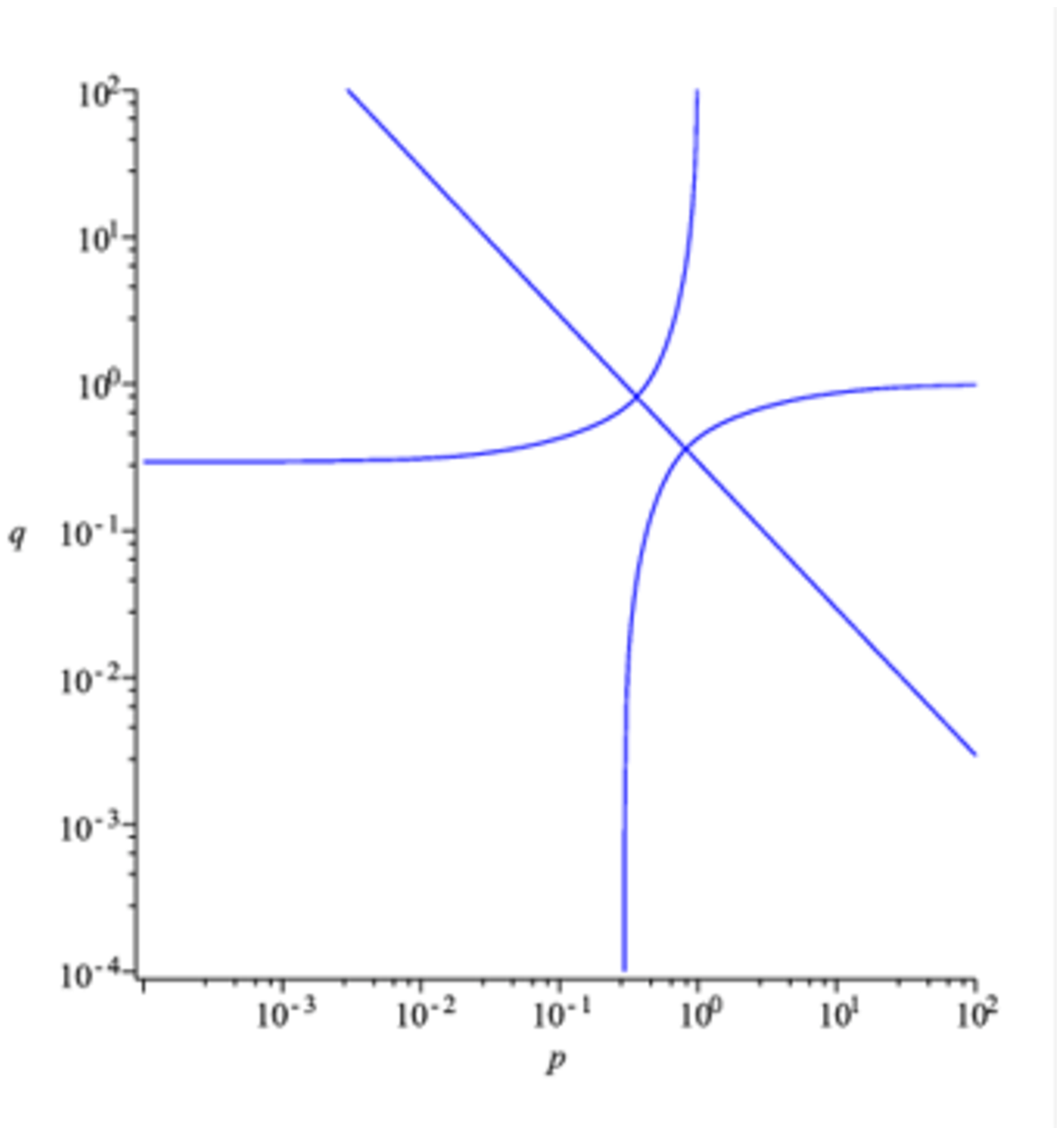}      %%[height=8cm, width=10cm]         
\caption{Crossover  of the curves (\ref{bq6}) having different topologies. Left
  panel: crossover from Type I. to Type II. for solitons moving in the same
  direction (at $k_1=1$, $k_2=1.57912575$), right panel: crossover from Type
  II. to Type $\overline{\rm II}$. for solitons moving in  opposite directions
  (at $k_1=1$, $k_2=1$).\label{f2}}          \end{center}          
\end{figure}

When changing the parameters $k_1$, $k_2$, the curves with different
topologies go over into each other. If one considers solitons moving into the
same direction, it is sufficient to consider the case $k_1<k_2$, since the
parameters must not coincide and the opposite case (i.e. $k_1>k_2$) simply
corresponds to the exchange of the axes. While the relative difference of
parameters is sufficiently large, we have the topology Type II. If the relative difference is diminished, we get the
topology Type I. The crossover between
the two topologies if shown in left panel of Fig.\ref{f2}. 

One might wonder
whether the middle branch of the curve touches both the other branches indeed at
the same parameters. The answer is affirmative, and follows again from the
symmetry (\ref{id1}). At the crossings of the branches one gets zero first
derivatives in two independent directions\footnote{This is actually an
  intersection of a
  saddle with the tangent plane.}, hence, one has simultaneously
\begin{eqnarray}
f(p,q,k_1,k_2)&=&0\;,\label{idc1}\\
\frac{\partial f(p,q,k_1,k_2)}{\partial p}&=&0\;,\label{idc2}\\
\frac{\partial f(p,q,k_1,k_2)}{\partial q}&=&0\;,\label{idc3}
\end{eqnarray}
        where
\begin{eqnarray}
f(p,q,k_1,k_2)=\sum_{n=1}^5\sum_{j=0}^n c_{nj}p^jq^{n-j}\;.\label{idc4}
\end{eqnarray}
Applying now the transformation (\ref{id1}) to a crossing point, it is
straightforward to show that at the transformed point
Eqs.(\ref{idc1})-(\ref{idc3}) are also satisfied. Indeed, the two crossings go
over into each other under transformation (\ref{id1})\footnote{The
  transformation is obviously an involution, i.e., it is equal to its own
  inverse.}. 

In case of oppositely moving solitons, $k_1=k_2$ is certainly possible, and a
transition from Type II. to Type $\overline{\rm II}$. occurs exactly when the
two parameters coincide (cf. right panel of Fig.\ref{f2}). 

\section{Time evolution of the extrema}

%Different ranges within the parameter space, corresponding to the three types of topologies  are shown in Figs.(\ref{f6})), (\ref{f7})).

\begin{figure}[H]                                                                 \begin{center}                                                                  \includegraphics[height=6cm]{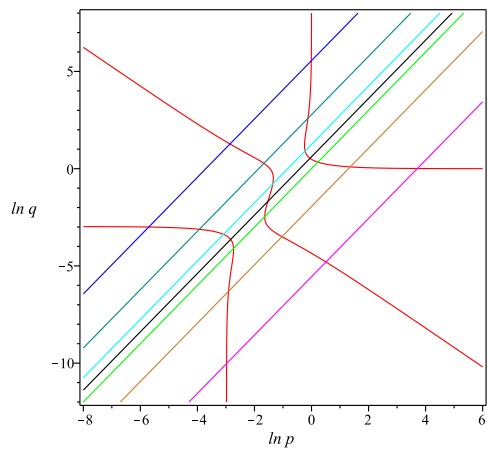}      %%[height=8cm, width=10cm]         
\caption{Type I. topology for intersections of the curve (\ref{bq6}) and (\ref{bq7}) at parameters $k_1=1.0$ and $k_2=1.5$ for solitons moving in the same direction. Different colors of the straight lines correspond to different time instants.\label{f8a}}      \end{center}          
\end{figure}

\begin{figure}[H]                                                                  \begin{center}                                                                  \includegraphics[height=6cm]{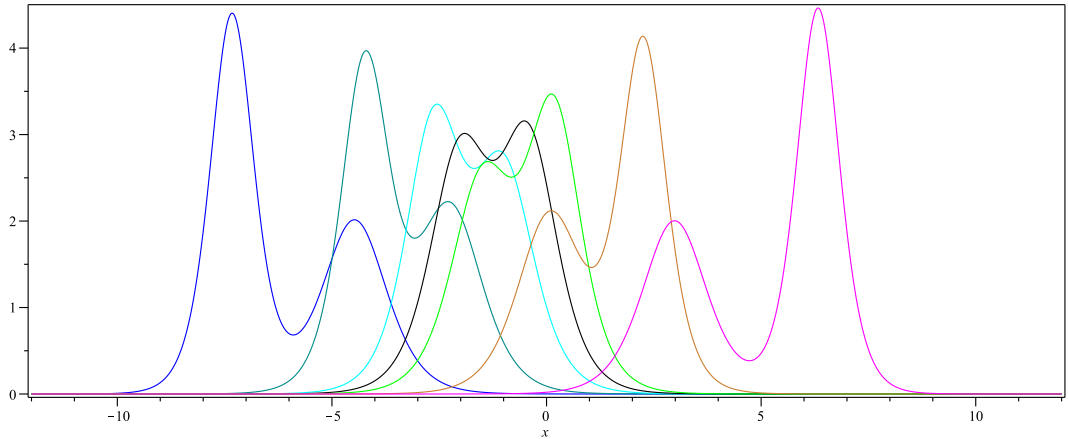}      %%[height=8cm, width=10cm]         
\caption{Type I. topology as seen in wave pattern for solitons moving in the same direction. Parameters and time instants identified by the colors are the same as in Fig.\ref{f8a}.      \label{f8b}}      \end{center}          
\end{figure}

Extrema of the waves are the intersection points of Eqs.(\ref{bq6}) and
(\ref{bq7}). As time goes on, the coefficient of the power function
(\ref{bq7}) changes from zero to infinity, and the corresponding
curve\footnote{On the log-log plots the graph of Eq.(\ref{bq7}) is a straight
  line.} ``sweeps through'' the curve (\ref{bq6}). According to the possible
shapes of curve (\ref{bq6}) we observe different scenarios, according to the
number of maxima. As shown in Figs.\ref{f8a} and \ref{f8b}, in the Type I. case we
always get three intersections, i.e.,
there are always two maxima and a minimum between them. 

\begin{figure}[H]                                                                  \begin{center}                                                                  \includegraphics[height=6cm]{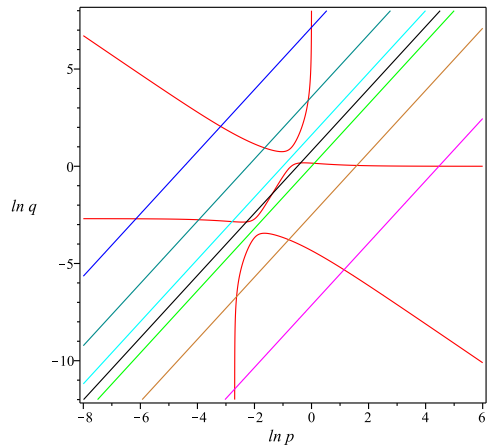}      %%[height=8cm, width=10cm]         
\caption{Type II.a topology for intersections of the curve (\ref{bq6}) and (\ref{bq7}) at parameters $k_1=1.0$ and $k_2=1.6$ for solitons moving in the same direction. Different colors of the straight lines correspond to different time instants.    \label{f9a}}      \end{center}          
\end{figure}

\begin{figure}[H]                                                                  \begin{center}                                                                  \includegraphics[height=6cm]{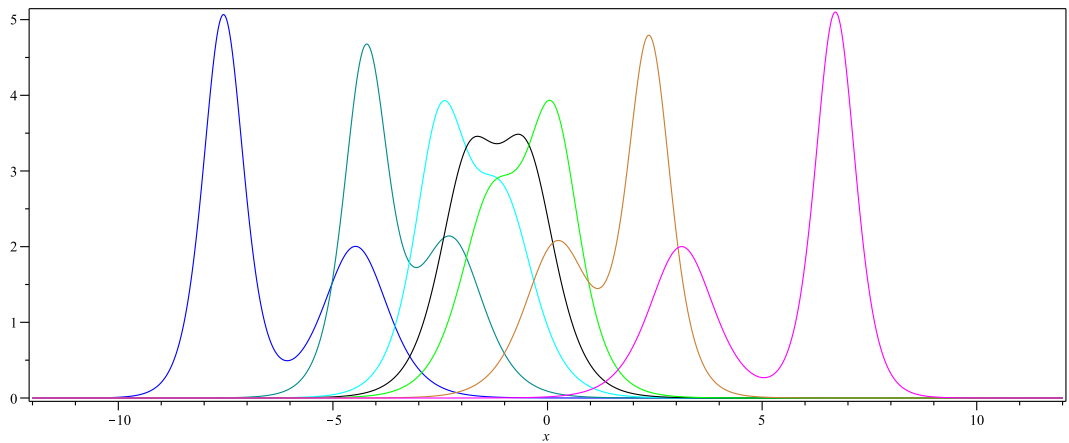}      %%[height=8cm, width=10cm]         
\caption{Type II.a topology as seen in wave pattern for solitons moving in the same direction. Parameters and time instants identified by the colors are the same as in Fig.\ref{f9a}.    \label{f9b}}      \end{center}          
\end{figure}

In contrast, in the
Type II. and Type $\overline{\rm II}$. cases there are periods when only a
single maximum exists. However, one has to distinguish here two subcases:

a) If the middle branch is steep enough, namely, if
\begin{eqnarray}
\frac{\partial \ln q}{\partial \ln p}>\frac{k_2}{k_1}\label{sqp}
\end{eqnarray}
at the symmetry point (\ref{sp}),
then at certain times the two maxima
reappear (Figs.\ref{f9a} and \ref{f9b}). 
This situation will be called Type
II.a (or Type $\overline{\rm II}$.a).

\begin{figure}[H]                                                                  \begin{center}                                                                  \includegraphics[height=6cm]{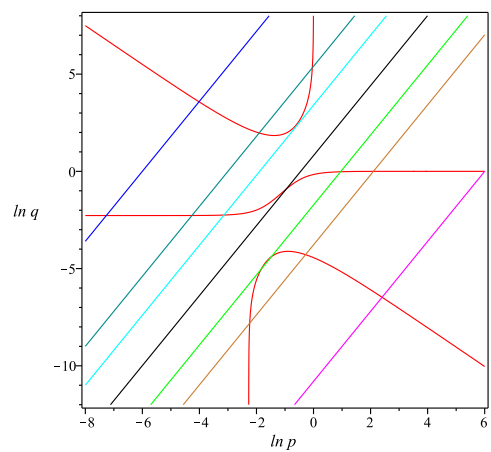}      %%[height=8cm, width=10cm]         
\caption{Type II.b topology for intersections of the curve (\ref{bq6}) and (\ref{bq7}) at parameters $k_1=1.0$ and $k_2=1.8$ for solitons moving in the same direction. Different colors of the straight lines correspond to different time instants.         \label{f10a}}      \end{center}          
\end{figure}

\begin{figure}[H]                                                                 \begin{center}                                                                  \includegraphics[height=6cm]{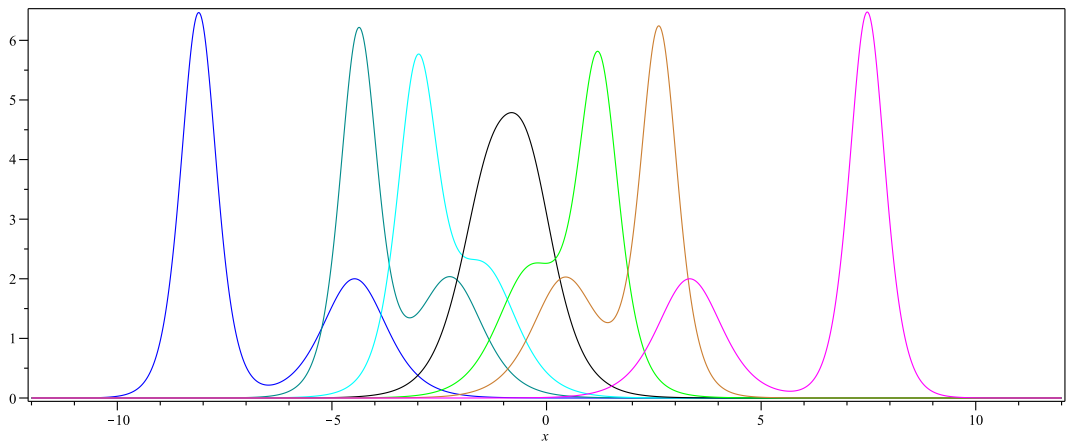}      %%[height=8cm, width=10cm]         
\caption{Type II.b topology as seen in wave pattern for solitons moving in the same direction. Parameters and time instants identified by the colors are the same as in Fig.\ref{f10a}.    \label{f10b}}      \end{center}          
\end{figure}

In such a case the initially well separated solitons (two maxima)
coalesc, the remnant of the smaller soliton being only a ``drooping shoulder'' at the
front side. Further
on, the ``shoulder'' moves towards the maximum, it rises and develops a second
maximum. Thus a shallow valley is created on the top of
the wave. At later times these events take place in reversed order: the rear
bank of the valley goes down, the corresponding maximum disappears and becomes
a drooping shoulder at the rear side, then it develops a maximum again behind the
taller wave and the two solitons are again separated.  

b) If the steepness of middle branch is smaller than $k_2/k_1$, no valley is
created on the top of the wave, as shown in Figs. \ref{f10a} and
\ref{f10b}. We shall call this case Type II.b (or Type $\overline{\rm II}$.b).

Both Type II.a and Type II.b can be observed in case of solitons moving in the
same direction. In contrast, for oppositely moving solitons, only the case Type II.b
(or Type $\overline{\rm II}$.b) can exist. This can be proven on the basis of
Eqs.(\ref{sqp}), (\ref{sp}). Indeed, we have at the symmetry point
\begin{eqnarray}
\frac{\partial \ln q}{\partial \ln p}=-\frac{4k_1^3(1-a)-a(k_2-k_1)^3+\sqrt{1-a}(4k_1^3-a(k_1+k_2)^3)}{4k_2^3(1-a)+a(k_2-k_1)^3+\sqrt{1-a}(4k_2^3-a(k_1+k_2)^3)}\;, \label{sqp1}
\end{eqnarray}
which is smaller than $k_2/k_1$ for oppositely moving solitons, if $k_2\ge k_1$.

\begin{figure}[H]                                                                 \begin{center}                                                                  \includegraphics[height=6cm]{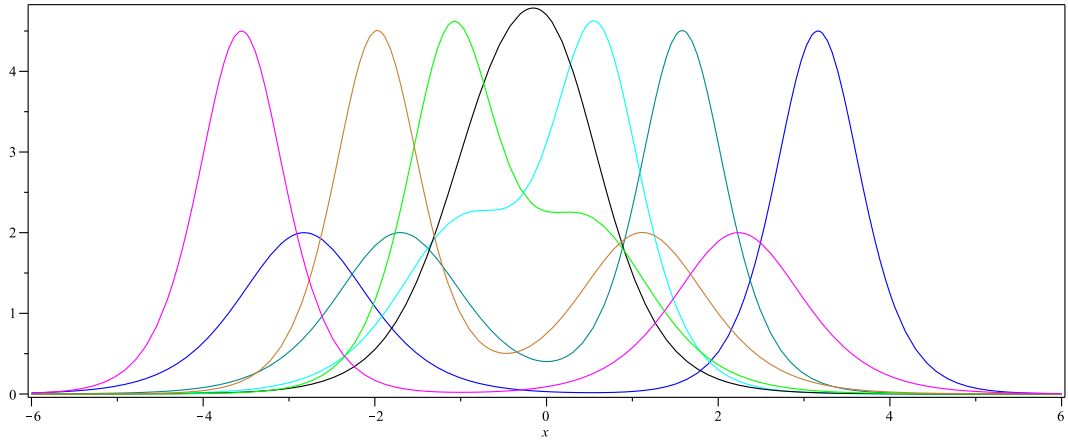}      %%[height=8cm, width=10cm]         
\caption{Type II.b topology as seen in wave pattern for solitons moving in opposite directions. Parameters and time instants identified by the colors are the same as in Fig.\ref{f11a}.      \label{f11b}}      \end{center}          
\end{figure}

\section{Parameter space}

\begin{figure}[H]                                                                  \begin{center}                                                                  \includegraphics[height=6cm]{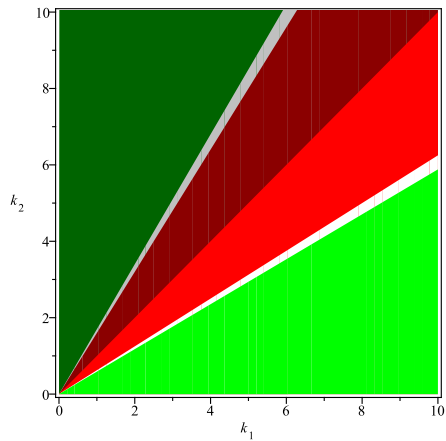}      %%[height=8cm, width=10cm]         
\caption{Scenarios shown in parameter space for solitons moving in the same direction. Legend: red, dark red: Type I. (see Fig.\ref{f8a} and Fig.\ref{f8b}), white: Type II.a (see Fig.\ref{f9a} and Fig.\ref{f9b}), gray: Type $\overline{\rm II}$.a, green: Type II.b (see Fig.\ref{f10a} and Fig.\ref{f10b}), dark green: Type $\overline{\rm II}$.b.      \label{f14}}      \end{center}          
\end{figure}

\begin{figure}[H]                                                                  \begin{center}                                                                  \includegraphics[height=6cm]{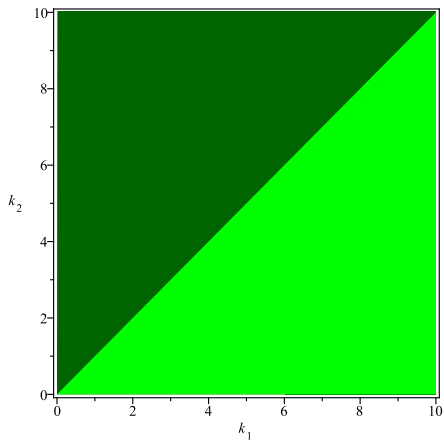}      %%[height=8cm, width=10cm]         
\caption{Scenarios shown in parameter space for solitons moving in opposite directions. Legend: green: Type II.b (see Fig.\ref{f11a}) and Fig.\ref{f11b} ), dark green: Type $\overline{\rm II}$.b.      \label{f15}}      \end{center}          
\end{figure}

The scenarios described above are summarized in parameter space in Figs.\ref{f14} and \ref{f15}. 

For solitons moving in the same direction, at the border between
Type II.a and Type II.b in parameter space  (see Fig. \ref{f14}.) the expression (\ref{sqp1}) is equal
to $k_2/k_1$.  This condition defines the border. As noted after Eq.(\ref{bq5}), for both $k_1\;,\;k_2\ll 1$ and  $k_1\;,\;k_2\gg 1$ the expression (\ref{sqp1}) depends only on the ratio $k_2/k_1$, hence the border looks linear. In fact, its slope slightly differs for small and large $k$ values.

As discussed above, the border between  between
Type I. and Type II.a in parameter space is given by Eqs.(\ref{idc1})-(\ref{idc4}). Again, the border is not exactly a straight line, its slope is slightly different for large and small $k$ values.

The shaded and the unshaded regions in Fig.\ref{f14}. are obtained by exchanging $k_1$ and $k_2$. 

As noted before, the $k_1=k_2$ line is not allowed.   

For solitons moving in opposite directions, the parameter space is even simpler (see Fig.\ref{f15}.). All parameter values are allowed, and the crossover from Type II.b to Type $\overline{\rm II}$.b occurs at $k_1=k_2$. 

%\begin{figure}                                                                  \begin{center}                                                                  \includegraphics[width=15cm]{probafuzes}      %%[height=8cm, width=10cm]         
%\caption{Left panel: Eq.(\ref{bq6}) for solitons moving in the same directions ($k_1=1$, $k_2=1.59$). Straight lines correspond to constant times $-2.0$, $-1.0$, $-0.5$, $-0.22$, $0.5$, $0.7$, $2.0$ (from top to bottom). Right panel:  the corresponding wave shapes, identified by their color.  Note the intermediate stages when only one maximum exist (curves in cyan and green).    \label{f6}}      \end{center}          
%\end{figure}   

%\begin{figure}                                                                  \begin{center}                                                                  \includegraphics[width=15cm]{probafuzes_00}      %%[height=8cm, width=10cm]         
%\caption{Left panel: Eq.(\ref{bq6}) for solitons moving in the same directions ($k_1=1$, $k_2=1.5$). Straight lines correspond to constant times $-2.0$, $-1.0$, $-0.5$, $-0.22$, $0.5$, $0.7$, $2.0$ (from top to bottom). Right panel:  the corresponding wave shapes, identified by their color. In contrast to Fig.(\ref{f6}), there exist always two maxima.     \label{f7}}      \end{center}          
%\end{figure}

\section{Nearly identical solitons}

\begin{figure}[htb]                                                                  \begin{center}                                                                  \includegraphics[height=6cm]{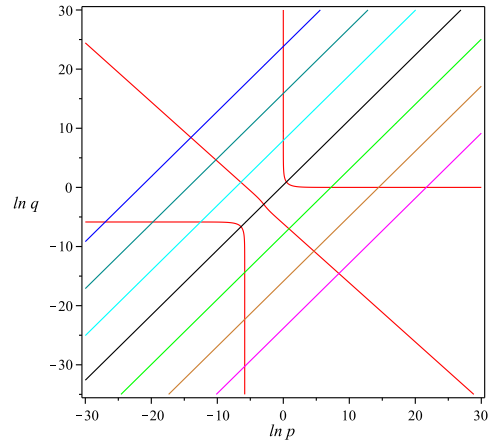}      %%[height=8cm, width=10cm]         
\caption{Intersections of the curve (\ref{bq6}) and (\ref{bq7}) at parameters $k_1=1.0$ and $k_2=1.1$ for solitons moving in the same direction.     \label{f13a}}      \end{center}          
\end{figure}

Let us consider now the situation when two nearly identical solitons interact. If they were strictly identical, we would obtain a single soliton solution rather than a two-soliton solution. Hence approaching the limit can be interesting. This situation corresponds to case I. Maxima correspond to intersections of the line (\ref{bq7}) with the two outer segments of the graph of (\ref{bq6}). Obviously, as time goes on, the initially well separated solitons  approach each other, then, without coalescing, their distance grows again. The minimal distance between them  may be estimated (cf. Fig.(\ref{f13a})) as

\begin{equation} 
\delta x_{min}\ge -\frac{1}{k_1+k_2}\ln(1-a)=\frac{1}{k_1+k_2}\ln\left(\frac{(k_1+k_2)^2+\frac{1}{12}\left(\frac{\omega_1}{k_1}-\frac{\omega_2}{k_2}\right)^2}{(k_1-k_2)^2-\frac{1}{12}\left(\frac{\omega_1}{k_1}-\frac{\omega_2}{k_2}\right)^2}\right).
\end{equation}

Evidently, $\delta x_{min}$ diverges logarithmically as $k_1\rightarrow k_2$.

\begin{figure}[htb]                                                                 \begin{center}                                                                  \includegraphics[height=6cm]{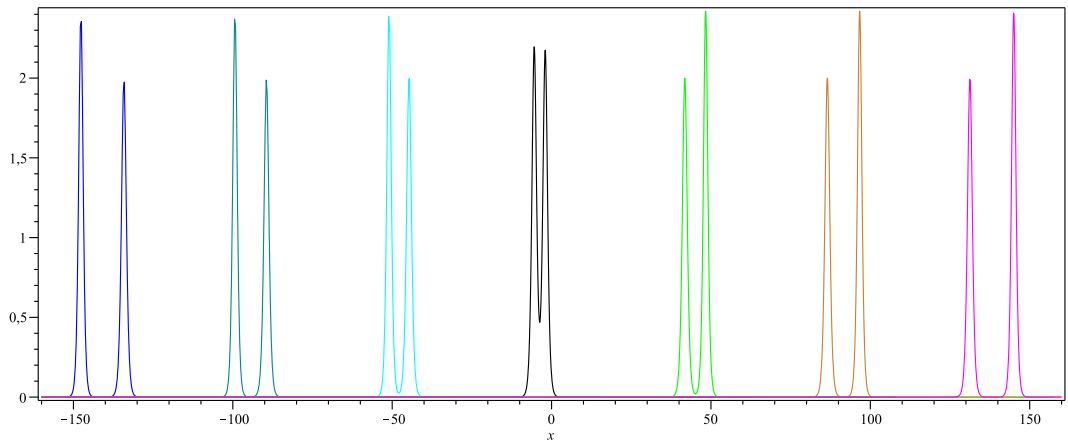}      %%[height=8cm, width=10cm]         
\caption{Wave patterns at consecutive time instants for parameters $k_1=1.0$ and $k_2=1.1$. The solitons are moving in the same direction. \label{f13b}}      \end{center}          
\end{figure}

If one considers now the corresponding waves (Fig.(\ref{f13b})), initially one sees two very similar solitons, the (slightly) taller one chasing the smaller one. When approaching, the tall soliton gradually loses its height and speed, at the same time, the smaller soliton gains hight and speed, the distance between them starts increasing, and eventually, we see the original solitons again, but this time the smaller one chasing and the taller one escaping.    

\section{Summary and discussion}
\begin{figure}[H]                                                                 \begin{center}                                                                  \includegraphics[height=6cm]{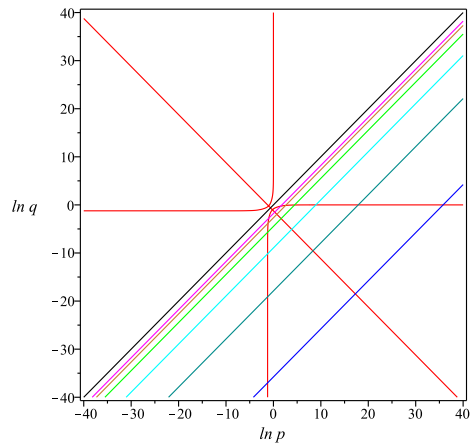}      %%[height=8cm, width=10cm]         
\caption{Intersections of the curve (\ref{bq6}) and (\ref{bq7}) at identical parameters $k_1=k_2=1.0$ for solitons moving in opposite directions.     \label{f12a}}      \end{center}          
\end{figure}

\begin{figure}[H]                                                                  \begin{center}                                                                  \includegraphics[height=6cm]{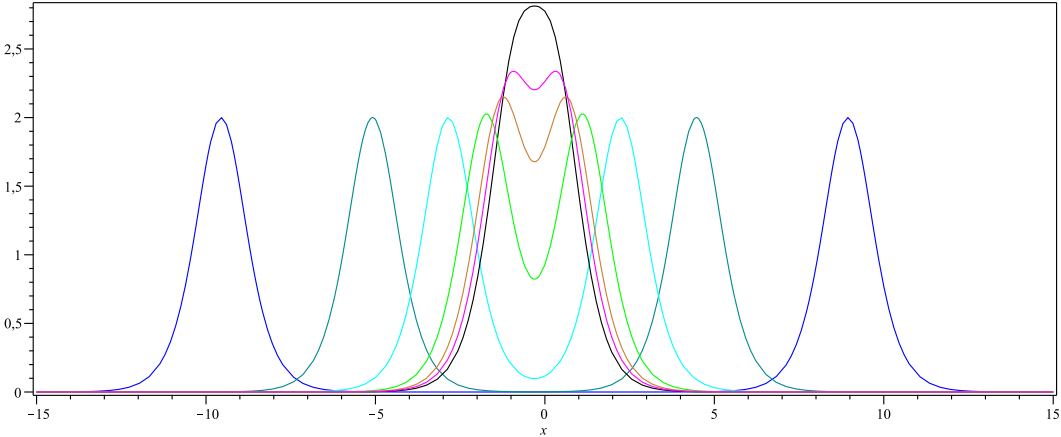}      %%[height=8cm, width=10cm]         
\caption{Wave patterns at consecutive time instants for identical parameters $k_1=k_2=1.0$. The solitons are moving in opposite directions.    \label{f12b}}      \end{center}          
\end{figure}

A simple classification scheme of the two soliton solutions of the Boussinesq equation have been presented. The scheme is based on the behavior of local maxima of the wave. We have shown that for solitons moving in the same direction there can be three different scenarios. In the Type I. case (see Figs. \ref{f8a} and \ref{f8b}) there are two maxima all the time, separated by a minimum. In the Type IIa. case (see Figs. \ref{f9a} and \ref{f9b}) initially, when the  solitons are still separated, there are two maxima with a minimum in between. During the collision the two solitons merge and only one maximum remains, the remnant of the other shows up only as a shoulder. Later on, however, the second maximum reappears and grows. Then the first maximum disappears for a while, but as the solitons become separated, it reappears and we have the initial solitons in a reversed ordering along the line. In the Type IIb. (see Figs. \ref{f10a} and \ref{f10b}) case the separated solitons merge to a wave having a single maximum, and later on a second maximum reappears and the solitons separate again. For solitons moving in opposite directions only the Type IIb. case exists (see Figs. \ref{f11a} and \ref{f11b}). In that case the wave numbers $k_1$ and $k_2$ may coincide (see Figs. \ref{f12a} and \ref{f12b}). In contrast, for solitons moving in the same direction $k_1$ and $k_2$ must be different. If $k_1\rightarrow k_2$ (see Figs. \ref{f13a} and \ref{f13b}) we have an extreme Type I. case, namely, there remains a large minimal distance between the solitons all the time, and the chasing soliton loses in height, while the escaping soliton gains in height during the collision. As a result, eventually the two solitons change their ordering along the line, without any close contact. 

Since our result are based on the long wave approximation, we expect that they should be observable in that limit. Also, the results for solitons moving in the same direction should follow from the Korteweg-de Vries equation as well, since that equation is obtained in the same approximation.  

\end{document}